\documentstyle[times,graphics,epsfig,amsmath,amssymb,natbib,referee]{mn2e}
\def\V{{\mathcal{V}}}
\def\S{\mathcal{S}}

\def\un{(\U,\nu)}
\def\tn{(\vec{\theta}, \nu)}
\def\U{\vec{U}}
\def\n{\hat{n}}
\def\t{\vec{\theta}}
\def\k{\hat{k}}

\def\mnras{MNRAS}

\def\aap{A\&A}

\newcommand{\be}{\begin{equation}}
\newcommand{\e}{\end{equation}}
\newcommand{\bear}{\begin{eqnarray}}
\newcommand{\ear}{\end{eqnarray}}

\def\mnras{MNRAS}
\def\prd{PRD}

\begin{document}

\title{The Effect of  $w-term$ on Visibility Correlation and Power
    Spectrum Estimation}
\author[Prasun Dutta, Tapomoy Guha Sarkar and S. Pratik Khastgir]
{Prasun Dutta$^{1}$\thanks{Email:
    prasun@cts.iitkgp.ernet.in},  
Tapomoy Guha Sarkar$^{2}$\thanks{Email: tapomoy@cts.iitkgp.ernet.in}
 and S. Pratik Khastgir
$^{1}$\thanks{Email: pratik@phy.iitkgp.ernet.in}
\\$^{1}$ Department of Physics and Meteorology \&
 Centre for Theoretical Studies, IIT Kharagpur, Pin: 721 302, India, 
\\$^{2}$ Centre for Theoretical Studies, IIT Kharagpur, Pin: 721 302, India.} 
\maketitle
\date{\today}

\begin{abstract}
Visibility-visibility correlation has been proposed as a technique for
the estimation of power spectrum, and used extensively for
small field of view observations, where the effect of $w-term$ is usually
ignored. We consider power spectrum estimation from the large field of
view observations, where the $w-term$ can have a significant
effect. Our investigation shows that a nonzero $w$ manifests itself as
a modification of the primary aperture function of the
instrument. Using a gaussian primary beam, we show that the modified
aperture is an oscillating function with a gaussian envelope. We show
that the two visibility correlation reproduces the power spectrum
beyond a certain baseline given by the width, $U_{w}$ of the modified
aperture. Further, for a given interferometer, the maximum
$U_{w}$ remains independent of the frequencies of observation. This
suggests that, the incorporation of large field of view in radio
interferometric observation has a greater effect for larger observing
wavelengths.   

\end{abstract}


\section{Introduction}
The directly observed quantity  in radio interferometry is the complex
visibility, measured as a function of baseline and frequency. Apart
from being the building blocks for radio imaging \citep{PSB89},  the observed
visibility data can be used for the estimation of various statistical
properties of the radio signal.
Visibility based estimators have been  widely used to estimate the  
power spectrum of intensity fluctuations of the observed
radio signal \citep{BS01,BA05,DBBC09a} and have also been
proposed to be a viable probe of Bispectrum \citep{ABP06,DSSA10}. For the Cosmic
Microwave Background Radiation (CMBR) 
signal, the formalism \cite{HM02} shows the effectiveness of using Maximum
Likelihood Estimators (MLE)  in
the Visibility space for direct estimation of the angular power
spectrum.
A large number of radio telescopes, like the presently functioning
GMRT \footnote{http://www.gmrt.ncra.tifr.res.in/}, upcoming
MWA \footnote{http://www.mwatelescope.org/} and SKA \footnote{http://www.skatelescope.org/} aim to
probe the universe 
through the redshifted $21$ cm observations. The HI power spectrum
is an unique observational probe of the dark ages \citep{BS01}, the epoch
of reionization \citep{DCB07} and the era, post-reionization
\citep{BSS09,ABP06}. The visibility based estimator  has 
advantages over the image based methods for the  estimation of power
spectrum. Apart from the fact that it uses the data in its raw form,
visibility based 
method is naturally more useful in cases of incomplete/sparse u-v
coverage. The method has also been used on smaller galactic scales for
estimation of HI  power spectrum of a turbulent ISM \citep{DBBC08,
  DBBC09a, DBBC09b} and
continuum power spectrum of Supernova remnant \citep{RBDC08}. 

The definition of visibility $\mathcal{V}$ is often approximated  as a
2D Fourier 
transform of the sky intensity distribution i.e,
$ {\mathcal V} ( u , v)  \stackrel{FT}{\rightleftharpoons}I ( l , m) $, where $
(l, m) $ denote 
the two angular coordinates  in the  sky and $ ( u, v )$,  the
corresponding  Fourier conjugate variables. While, this is true for
small field of view, and co-planar array distribution, the actual
3D visibility as is observed in radio interferometers,
do not satisfy this simple relationship. 
For most practical purposes, at high observing frequencies,
this approximation works reasonably well for the existing radio 
interferometers like the GMRT , VLA
etc. However, upcoming radio telescopes aiming at wide field 
imaging will require inclusion of the full 3D effect.
The same problem is also faced in imaging for large
field of view observations or observations with non-coplanar baselines and  is
tackled by techniques of mosaicing \citep{PSB89, FD80, CP92, CGB08}. 

However the problem of
estimating power spectra using the measured three dimensional
visibility $ {\mathcal V}(u,v,w) $, has not been addressed. 

In this letter we investigate  the effect of using the three dimensional
visibility in the estimation of power spectrum and hence find a
possible justification of the  simpler two dimensional
approach used extensively earlier.
\section{Effect of   \lowercase{$w-term$}  on the Aperture Function}

The direct observable  in the radio interferometric observations is the
complex visibility $\V^{3D} \un$. For a pair of antennae  separated by
$\vec{d}$, with
  each antenna pointing along the direction of the unit vector $\hat{k}$ 
(referred to as the phase center) we have 
\begin{equation}
{\V}^{3D} \un = \int d \Omega_{\hat{n}}  e^{2 \pi  \U \cdot
(\n -\k)}  A(\n - \k,\nu)   I(\n - \k,\nu)  
\end{equation}
where $\n$ denotes the unit vector to different directions of the sky,
baseline  $\U = {\vec{d}} /\lambda $,  $A(\n - \k,\nu)$  denotes the
primary beam and $I(\n - \k,\nu)$ is  the specific intensity.
Writing $\U = \U_{\perp} + w\k$, where $\U_{\perp}$ is a 2D vector, 
and defining $ \n - \k = \vec{\theta} $,  we have,  for
$ |\vec{\theta}| \ll   1$, \  $\vec{\theta}\cdot \k = 0 $, implying that 
$\vec{\theta}$ is a $2D$  vector. In this limit $\vec {\theta}$ gives
the position of any point  on the sky with
respect to the phase centre in a 2D tangent plane. This is known as
the flat-sky approximation. The term $ w\k $
quantifies deviation from this.

In the  $2D $ approximation we have
\be
{\V} ^{2D}(\U_{\perp},\nu) = \int d^2\t   \ e^{ 2 \pi i  \U_{\perp} \cdot
\t} \ A(\t,\nu) \  I(\t,\nu)
\e
Writing  the specific intensity $I\tn$ as $I\tn = \bar{I}_{\nu} +
\delta I\tn $,  where the first term is a  constant background and the
second term is a fluctuation,  we have
\be
{\V}^{2D}(\U_{\perp},\nu) =  \bar{I}_{\nu} \tilde{A}(\U_{\perp},\nu) + \tilde{A} (\U_{\perp},\nu)\otimes \tilde{\delta I}(\U_{\perp},\nu)
\e
where tilde represents a Fourier transform and $\otimes$ denotes a
convolution. 

The aperture function $\tilde{A}(\U_{\perp},\nu)$, peaks at
$\U_{\perp} = 0$ and has a finite width. Hence,
we shall retain the second term in all subsequent discussions. 

We note that, ignoring the
$w-term$ leads to a simplification of the expression for visibility
and in the 2D approximation, $ \V^{2D}(\U_{\perp},\nu)$ is the
Fourier Transform of $\ A(\t,\nu)  \delta I(\t,\nu)$.
Hence, we have,
\be
\ A(\t,\nu) \  \delta I(\t,\nu) = \int  d^2 \U_{\perp}'\  \V^{2D}(\U_{\perp}',\nu)  \ e^{-2 \pi i.
  \U_{\perp}' \cdot \t}   
\e
Substituting in equation (1) we obtain
\be
\V^{3D}\un = \int   d^2 \U_{\perp}'  \ K(\U, \U_{\perp}', \nu) \  \V^{2D}(\U_{\perp}',\nu)
\e
Where the kernel $ K(\U, \U_{\perp}', \nu)$ is given as
\begin{eqnarray}
K(\U, \U_{\perp}', \nu)&=&  \int d \Omega_{\hat{n}}\ e^{-2 \pi i \ 
  (\U_{\perp}' - \U) \cdot (\n -\k)} \nonumber \\
&=& 4 \pi j_{0}(2 \pi \mid \U_{\perp}' - \U \mid ),
\end{eqnarray}
with $j_{0}$ denoting the Spherical Bessel function.

Defining a  quantity $\tilde{\mathcal A}\un $ as
\be
\tilde{{\mathcal A}} \un = 4\pi \int d^2\U_{\perp}' \   j_{0}(2\pi|\U_{\perp}' - \U|)
\tilde{A}(\U_{\perp}',\nu). 
\label{eq:modap}
\e
The 3D visibility takes the form
\be 
\V^{3D}\un = \int d^2\U_{\perp}' \ \tilde{{\mathcal A}} (\U -
\U_{\perp}')  \ \tilde{\delta I}(\U_{\perp}')
\label{eq:V23D}
\e
It is to be noted that the `$w$' dependence of $\V^{3D}\un$ is
translated to the function $\tilde{{\mathcal A}} (\U -
\U_{\perp})$. This can be regarded as a modified aperture function. 
 We investigate the nature of the modified aperture $\tilde{{\mathcal
     A}}(\U_{\perp},w )$ as a
 function of $U_{\perp}$ at different values of $w$. 
We have assumed the primary aperture $ A( \U_{\perp}) $   to be a
gaussian, $\exp \left [ -\frac{U_{\perp}^{2}}{2 U_{0}^{2}} \right ]$,
of width $ U _{0}$, and evaluated the  integral in
 Eqn.~(\ref{eq:modap}) numerically. This indicates that
 $\tilde{{\mathcal A}} \un$ has an 
 implicit dependence on $U_{0}$. For an antenna of diameter $D$,
 $U_{0}$ can be approximately written as $U_{0} \sim D /\lambda$ for
 observing wavelength $\lambda$. We use $U_{0} = D /\lambda$ with
 $D=45$ m   (specifications of the GMRT) for the subsequent
 discussion. This corresponds to a field of view of $4^{\circ}$ at
 $150$ MHz. We shall discuss the effect of larger field of view later.
\begin{figure*}
\begin{center}
\mbox{\epsfig{file=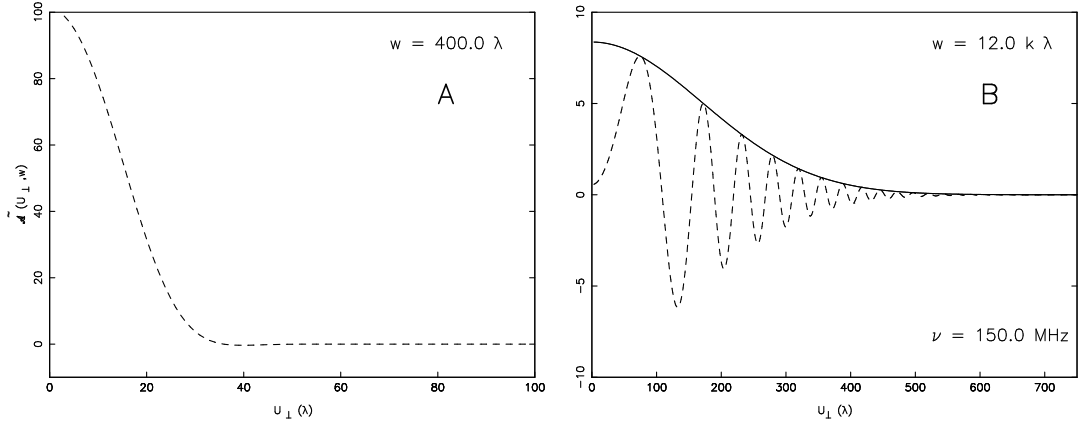,width=2.2in,angle=90}}
\end{center}
\caption{Modified aperture $\tilde {\mathcal A}(\U_{\perp},w )$
 plotted as a function of $\U_{\perp}$ for two different values of
  $w$, (A) $w=400. \lambda$ and (B) $w = 12.$ k$\lambda$ at $\nu
= 150$ MHz. Solid line in B shows the gaussian envelope.}
\label{fig:1502p}
\end{figure*}

{\bf Figure}~\ref{fig:1502p} shows the variation of  the modified 
aperture $ \tilde{{\mathcal A}} (\U_{\perp}, w) $ as a function of $ 
U_{\perp}$ for two representative values of $ w $ (A: $ w = 
400\ \lambda$ , B: $ w = 12\ {\rm k}\lambda$) at frequency $ \nu = 150 
\rm{MHz}$. 
\begin{figure}
\begin{center}
\mbox{\epsfig{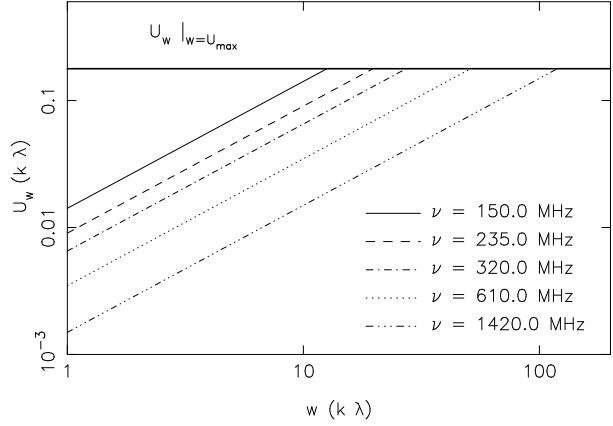}}
\end{center}
\caption{$U_{w}$ is plotted as a function of $w$ for different central
  frequencies of GMRT. Horizontal solid line corresponds to the
  maximum possible baseline.
}
\label{fig:Uw}
\end{figure}

\begin{figure}
\begin{center}
\mbox{\epsfig{file=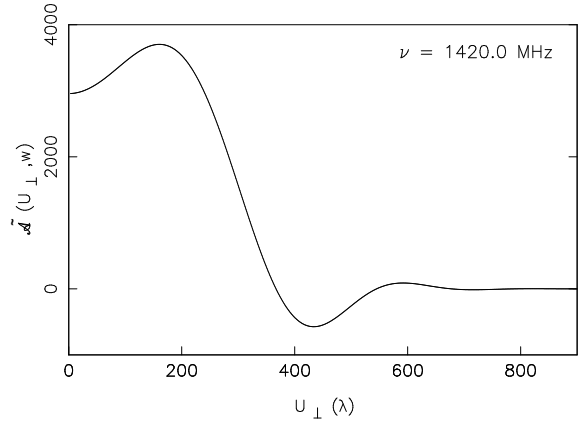,width=2.2in,angle=90}}
\end{center}
\caption{Modified aperture $\tilde {\mathcal A}(\U_{\perp},w )$
 plotted as a function of $\U_{\perp}$ for  $w=120.$ k $\lambda$ at
 $\nu=1420.$ MHz.
}
\label{fig:14201p}
\end{figure}
For $ w = 0 $ one reproduces the primary gaussian aperture
trivially. We also note that for $ w \ll U_{\perp}  $ the gaussian
profile is still maintained ({\bf Figure}~\ref{fig:1502p}:A). However for
large values of $w$ the 
aperture function manifest oscillations ({\bf Figure}~\ref{fig:1502p}:B).
The  period of these oscillations is found to be 
sensitive to $w$ (decreasing as $w$  increases).
The envelope of the modified aperture is also a  gaussian,
  $ C  \exp \left [ - \frac{U^2}{2 U_w^2} \right ] $, with 
the  parameter $U_w$ being a measure of it's dispersion.
Ignoring the effect of oscillations in $\tilde{{\mathcal A}}$, we note
that the 3D formalism can be recast in the same form as it's 2D   
counterpart with $U_{w}$ taking the role of $U_{0}$. 
\citet{HM02} have obtained a similar result 
 assuming small field of view, where, they have shown that the
 effect of $w$-distortion can be considered as turning the primary
 beam into a complex gaussian.  

We next investigate the $w$ and frequency dependence of $U_{w}$. For a
given frequency at large $w$, $U_{w}$ is found to increase linearly
with $w$, i.e, $U_{w} \sim m(\nu) w$.  This implies
that the $w-term$ effectively broadens the aperture of the
instrument. {\bf Figure}~\ref{fig:Uw} shows the variation of $U_{w}$ with
$w$ for different frequencies in log-log scale.  We note that, the
slope $m(\nu)$ (as represented by the y-intercept in the
{\bf Figure}~\ref{fig:Uw}) determines the  effect of $w-term$ for increasing 
values of $w$. A small value of $m(\nu)$ implies  a slow increase of
$U_{w}$ with $w$ and the effect of the $w-term$ is
less. $m(\nu)$ is found to fall off as $\sim 1/\nu$ with
frequency. This indicates that the departure from the 
flat-sky approximation is more pronounced at the lower
frequencies. Redshifted 21 cm line observed at frequency $\nu$ probes
the redshift $z =\left [ \frac{1420 ({\rm 
  MHz})}{\nu} -1 \right]$. Hence, one may expect the $w-term$ to have
a greater  effect while probing higher red-shifts. 

21 cm line has been used to study the ISM dynamics of the nearby galaxies
$(z \sim 0)$. {\bf Figure}~\ref{fig:14201p} shows the modified aperture
function  for frequency $1420$ MHz and $w = 120\ {\rm k}\lambda$
(this being the maximum $U$ for GMRT like arrays). At this frequency,
for GMRT, $U_{0} = 0.1  {\rm k}\lambda$, whereas $U_{w}\mid _{w =
  U_{max}} = 0.18  {\rm 
  k}\lambda$. It follows that the flat-sky approximation can be
safely used if the largest length scale probed, corresponds to a $U_{\perp} \gg
3\ U_{w}\mid _{w =U_{max}}$. 

Till now, we have investigated the effect of $w-term$ using $D=45$ m,
which corresponds to an field of view of $4 ^{\circ}$ at 150 MHz. We
estimated $U_{w}\mid _{w =U_{max}}$ assuming $D = 4$ m to $D=45$ m at
$\nu = 150$ MHz. At $D=4$ m,  (which corresponds to the largest
proposed field of view, $45^{\circ}$, of SKA), $U_{w}\mid _{w =U_{max}}
= 4$ k$\lambda$ for $U_{max} = 120$ k$\lambda$.  We have also 
observed that,  $U_{w}\mid _{w =U_{max}} \sim  1/D$. This indicates that the
effect of $w-term$ is more pronounced for larger field of view, as
expected. 

\section *{Visibility Correlation and Power Spectrum estimation}
The power spectrum, $ P(U_{\perp},\Delta\nu  ) $ of the intensity fluctuations $\delta I $ on the
sky is defined as
\be
\langle \tilde{\delta I}(\U_{\perp }, \nu_1) \ \tilde{\delta I}(
\U'_{\perp }, \nu_2)
\rangle = P(U_{\perp}, \Delta\nu) \  \delta^2_{\bf{D}}(\U_{\perp } -
\U'_{\perp })
\e
where  $ \Delta\nu = |\nu_1 - \nu_2| $.
We shall be considering, for simplicity,  $\nu_1 = \nu_2$ in all
subsequent discussions and hence drop the $ \Delta \nu$ dependence of
the power spectrum. 

We define
\be
V_2^{3D}(\U_{ a},  \U_{b}) = \langle
\V^{3D}(\U_{ a}) \  \V^{3D  *}(\U_{ b}) \rangle.
\e
Using Eqn.~(\ref{eq:V23D}) we obtain 
\begin{eqnarray}
V_2^{3D}(\U_{ a},  \U_{b}) = \int d^2\U_{\perp}^{''} \int d^2
\U_{\perp}^{'} \tilde{\mathcal{A}} (\U_{a} - \U_{\perp}^{''} ) \nonumber \\
\tilde{\mathcal{A}} ^*( \U_{ b} - \U_{\perp}^{'} )
\langle \tilde{\delta I}  (\U_{\perp}^{''} ) \ \tilde{\delta I}^*
(\U_{\perp}^{'} )\rangle 
\label{eq:delU}
\end{eqnarray}
Using the definition of the power spectrum and considering the
correlation at the  same base-line $\U$, this simplifies to 
\be
V_2^{3D}(U) = \int  d^2 \U_{\perp}^{'} \ {|\tilde{\mathcal{A}} (\U_{} - \U_{\perp}^{'})|}^2  \ 
 P(U_{\perp}^{'})
\label{eq:est}
\e
Noting that the effect of the $w-term$ is contained in the modified
aperture $\tilde {\mathcal A}$ we can retrieve the 2D estimator $V_2^{2D}$
used earlier \cite{DBBC09a} by replacing  $ \tilde{\mathcal A}$ with
$\tilde{A}$. Hence, 
\be
V_2^{2D}(U_{\perp})= \int  d^2 \U_{\perp}^{'} \ {|\tilde{A} (\U_{\perp} -
    \U_{\perp}^{'})|}^2  \  
 P(U_{\perp}^{'}).
\e
We shall now discuss the effect of the  $w-term$ on the estimator
defined in Eqn.~(\ref{eq:est}).

\begin{figure}
\begin{center}
\mbox{\epsfig{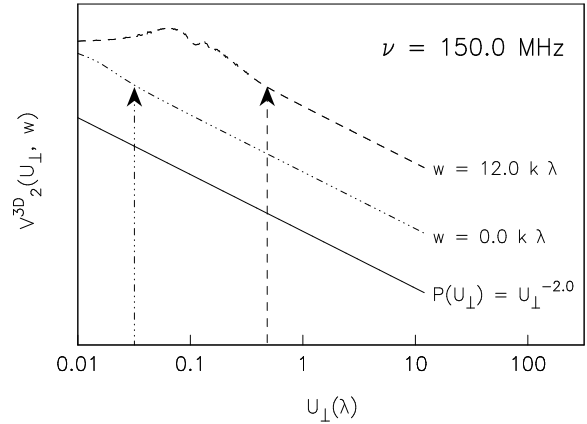}}
\end{center}
\caption{$V^{3D}_2$  as a function of $U_{\perp}$ for  $w=0$
  (dot-dash) and $w=U_{max}$ (dash) at $\nu = 150$ MHz, assuming $P(U_{\perp})
={U_{\perp}}^{-2}$. We also plot $P(U_{\perp})={U_{\perp}}^{-2}$
(solid line) for reference. The vertical arrows show the $U_{\perp}$
value above which the power law is recovered at $1 \%$. Note
that the plots are given arbitrary offset for clarity. 
}
\label{fig:pspec}
\end{figure}

{\bf Figure}~\ref{fig:pspec} shows $V^{3D}_2$  plotted
as a function of $U_{\perp}$ for two values of $w$, ($ w = 0 $ and $ w
= U_{max}$) at $ \nu = 150$ MHz, assuming $ P(U_{\perp})
={U_{\perp}}^{-2}  $. We have chosen $ U_{max}= 12 k\lambda$ (this
being  the largest baseline for the GMRT at $150 $ MHz). We  have
shown the  power 
spectrum $  P(U_{\perp})={U_{\perp}}^{-2}$ for comparison.
For  large values of  $w$, $V^{3D}_2$ show oscillations for $
U_{\perp} < U_w$,  which arises due to  the oscillatory nature of 
$ |\tilde{\mathcal{A}}(U)|^2$.

We find that  $ V^{3D}_{2}$
faithfully recovers the power law  $ {U_{\perp}}^{-2}$ (at $1\% $) for
$U_{\perp}$ greater than a certain value. This  value  is found to be
$3 \times(\sqrt{2}U_{0})$ for $ w = 0 $
and $ 3 \times(\sqrt{2}U_{w}) $ at $ w = 12\ {\rm k} \lambda$. Hence, a
non-zero $w-term$  changes the $U_{\perp}$
value beyond which the power spectrum estimation would be valid. 

The quantity of interest in power spectrum estimation using the the
radio interferometric observations used earlier \citep{BA05,DBBC09a}
is
\begin{equation}
{\mathcal E}(U_{\perp}) = \int_{0}^{U_{max}} dw \ \rho(w)\ V^{3D}_{2}(U_{\perp}, w),
\end{equation}
where, $\rho(w)$ is a normalized probability distribution of
$w$. The function $\rho(w)$ is specific to an observation as well
as to the array configuration of the interferometer. Hence, it is
difficult to make a general quantitative statement regarding the
effect of $w-term$ in $\mathcal{E}$.
 Since, for a given $w$, the largest baseline above which
$V^{3D}_{2}(U_{\perp}, w) \sim P(U_{\perp})$ is $U_{w}$,
 we can qualitatively state that ${\mathcal E}$ gives a good estimation
of the power spectrum for $U \ge 3\ U_{w} \mid_{ w  = U_{max}}$. It is
important to note that, for a specific array configuration, $U_{w}
\mid _{w = U_{max}}$ is independent of the frequency $\nu$, whereas $U_{max} \propto
\nu$ ({\bf Table}~\ref{table:t1}). Hence, the $U_{\perp}$ range amenable for power spectrum
estimation is larger for large observing frequencies.

\section{discussion and conclusion}
\begin{table}
\centering
\begin{tabular}{|c|c|c|}
\hline 
& $w = 0$ & $w \sim U_{max}$ \\
\hline
\hline
Aperture & $\tilde {A}(U_{\perp})$ & $\tilde{\mathcal A} (U_{\perp}, w)$ \\
&width $U_{0}$ & width $U_{w}$ \\
\hline
Visibility & & \\
correlation & $V_{2}^{2D}(U_{\perp})$ & $V_{2}^{3D}(U_{\perp}, w)$ \\ 
\hline 
\end{tabular}
\caption{The effect of $w-term$, comparison between various quantities.} 
\label{table:t1}
\end{table}

\begin{table}
\centering
\begin{tabular}{|c|c|c|c|}
\hline 
& & $150$ (M Hz) & $1420$ (M Hz)\\
\hline
\hline
$U_{0}$ & & 0.01 &  0.1 \\
$U_{max}$& & 12.0 & 120.0 \\
$U_{w}\mid _{w = U_{max}}$& & 0.18 & 0.18 \\
\hline 
\hline 
\end{tabular}
\caption{Relevant $U_{\perp}$ (k $\lambda$) values at different frequencies.} 
\label{table:t2}
\end{table}

We have studied the effectiveness of the widely used power spectrum
estimator in presence of  a non zero $w$. The $w-term$ is found to
affect the visibility correlation estimator through a modification of
the aperture function ({\bf Table}~\ref{table:t1}). The effect is more
pronounced for the lower frequencies, where, for a given
interferometer, the baseline range over 
which $V_{2}^{3D}$ reproduces the power spectrum, is reduced. This
restricts the largest possible length scales that can be probed using the $21$
cm radiation from the epoch of  reionization ($20 \lesssim z \lesssim
6$). However, for the observations of the nearby universe (i.e,
$\nu=1420$ MHz), $w-term$ effect is not too significant ({\bf
  Table}~\ref{table:t2}). 

The two-visibility correlation defined in Eqn.~(\ref{eq:est}) has a
positive noise bias associated with it, which can even exceed the signal
in radio interferometric observations. To reduce the effect of
the noise bias, one may follow the method used
in the 2D analysis \citep{BCB06,DBBC08,DBBC09a} considering correlation of
the  visibilities 
at two nearby baselines (Eqn.~(\ref{eq:delU}), with $\vec{U}_{b} =
\vec{U}_{a} + \Delta \vec{U}$). For simplicity, we have
considered visibility correlations at the same baseline
(Eqn.~(\ref{eq:est})). However, we note that an alternative method for reducing
the noise bias is to observe the same field repeatedly for the same
baseline configuration.

Redshifted 21 cm observations allow us to probe the HI distribution of the
universe over continuously varying redshifts by tuning the frequency
of the radio observations. By correlating visibilities at different
frequencies it is possible to do a tomographic study of the HI
distribution, thereby probing the 3D power spectrum. In this work we
have restricted ourselves to 
visibility correlations at the same frequency of observation. Future
investigations may consider the effect of a nonzero $\delta \nu$ on
power spectrum estimation.

\section*{Acknowledgments}
PD and TGS are thankful to Somnath Bharadwaj, Tirthankar Roy Chowdhury,
Sk. Saiyad Ali, Kanan Datta, Tatan Ghosh, Suman Majumder, Abhik Ghosh,
Subhasis Panda and  Prakash Sarkar for use full
discussions. PD would like to acknowledge SRIC, IIT, Kharagpur for
providing financial support. TGS would like to acknowledge financial
support from BRNS, DAE through the project 2007/37/11/BRNS/357.

\end{document}